\documentclass[12pt]{article}
\usepackage{amsmath}
\usepackage{graphicx}
\usepackage{hyperref}
\usepackage{bibentry}
\usepackage[latin1]{inputenc}

\title{Fast Bitmap Fit: A CPU Cache Line friendly memory allocator for single object allocations}
\author{Dhruv Matani, Gaurav Menghani}
\date{\today}

\begin{document}
\maketitle

\begin{abstract}

Applications making excessive use of single-object based data structures (such as linked lists, trees, etc...) can see a drop in efficiency over a period of time due to the randomization of nodes in memory. This slow down is due to the ineffective use of the CPU's L1/L2 cache. We present a novel approach for mitigating this by presenting the design of a single-object memory allocator that preserves memory locality across randomly ordered memory allocations and deallocations.

\end{abstract}

\section{Introduction}

Numerous applications make use of data structures that allocate single objects (i.e. objects of a fixed size). For example linked lists (singly/doubly linked), trees (balanced/unbalanced), etc... all allocate nodes of a fixed size. These nodes contain both metadata (pointers to other nodes) as well as data (the actual values being stored).

Over the course of the application's lifetime, the application may create, update, and delete such data structures. Sorting or reversing a linked list changes the order of these nodes, and balanced trees typically move nodes around quite a bit.

Allocation algorithms that are commonly used on these systems (e.g. first fit\cite{firstfit}, best fit, cached, etc...) optimized for allocation/deallocation speed, and may keep these nodes around in a random order (or the order in which they were returned). This problem is especially exacerbated in allocators that implement arenas and keep small allocated objects around for immediate future use\cite{dirk93}. The next time memory is requested, it comes back in a random order, and even a simple operation such as a linear traversal of a newly created linked list can be much slower when the memory returned by the allocator is in a random order. This is because the memory could be in different cache lines and may not all fit in the CPU's L1/L2 cache. Furthermore, if the object size itself is small, then the CPU's L1/L2 cache may be wasted by pulling in memory that is entirely unused.

Previous research on using a software look-aside buffer has been done to mitigate these effects for linked lists\cite{gerald84}. However, these approaches are complex, introduce implementation overhead, and don't scale to all single-object based data structures.

We discuss a simple and elegant approach for an efficient single object memory allocator which ensures that repeated allocation/deallocation cycles of memory in random order don't affect future allocations, which will be serviced in a cache friendly manner. Further, this structure can be exploited by the application by requesting memory that is "close" to an already allocated memory block. This is beneficial if the application expects both nodes in the data structure to be needed together. For example, if the user wishes to insert an element into an existing linked list in the middle, it can pass in the address of the previous (or next) node (hint) in the list and the allocator should attempt to service this request by returning a memory address (node) as close as the provided "hint".

\subsection{Pseudo-code highlighting a problematic case}

The pseudo-code below shows an example of how memory is allocated, used, deallocated, and then allocated again resulting in poor memory locality and cache performance for a simple data structure such as a linked list.

\begin{verbatim}
List lst = {}; // empty list.

// Populate the list.
for (i = 0; i < 10000; ++i) {
  // Insert a random number in the range [0..100] into the list.
  lst.append(random(100));
}

// Iterate over all the elements of the list.
// it is an iterator into the linked list lst.
for (it = lst.begin(); it != lst.end(); ++it) {
  // This will access memory in the order returned by the
  // allocator. Typically, for the first use, this is in address
  // order.
  print(it.value());
}

// Sort the list in ascending order of element value.
lst.sort();

// Return memory for linked list nodes back to the allocator.
lst.clear();

// Populate the list again.
for (i = 0; i < 10000; ++i) {
  // Insert a random number in the range [0..100] into the list.
  // This time around, the memory returned by the allocator
  // could be either in memory order or in the order returned
  // by List::clear(). If it's the latter, then consecutive nodes
  // in the linked list will be randomly ordered.
  lst.append(random(100));
}

// Iterate over all the elements of the list.
// it is an iterator into the linked list lst.
for (it = lst.begin(); it != lst.end(); ++it) {
  // This will access memory in the order returned by the
  // allocator. If the memory address of the nodes is random,
  // this will thrash the cache, and negatively impact the
  // application's running time due to an increase in CPU
  // cache misses.
  print(it.value());
}
\end{verbatim}

\subsection{The impact of memory locality}

Modern CPUs have L1/L2 caches to make references to main memory faster for data that can be stored in the CPU cache. Access to data in the L1 and L2 caches can be 200x and 15x faster than access to main memory\cite{ladis2009}.

\begin{center}
\begin{tabular}{ |c|c|c| } 
 \hline
 Access Type & Latency (ns) & Relative to Main Memory Reference \\ 
\hline
 L1 cache reference & 0.5 ns & 200x \\ 
 L2 cache reference & 7 ns & 15x \\ 
 Main memory reference & 100 ns & 1x \\ 
 \hline
\end{tabular}
\end{center}

Hence, it's very much preferred to keep as much of the active data set in the CPU's L1/L2 cache. Once data is in the cache, every attempt should be made to use as much of it as possible. Alternatively, we should attempt to lay the data out in memory such that data used together in time is laid out close together in space (in memory).

Additionally, application inefficiencies resulting from inefficient placement of data are very hard to detect and reason about \cite{dirk93,yehuda2011}. Hence, every effort must be made to ensure that programming abstractions support effective data locality by design.

\section{Previous Work}

\subsection{General Purpose Memory Allocators}

General purpose memory allocators have been built and studied for a long time. There are some popular algorithms that are used very commonly.

\begin{enumerate}

\item First Fit\cite{knuth73}: Typically implemented to optimize for allocation speed at the expense of memory fragmentation.
\item Best Fit\cite{paul95}: Typically implemented to minimize memory fragmentation at the expense of allocation speed.
\item Binary Buddy\cite{knuth73}: Memory returned in sizes that are the closest power of 2 greater than or equal to the requested memory size.
\item Bitmap Fit\cite{brian08}: Bitmaps are used to keep track of fixed-size memory blocks that are free/used. Bitmaps need to be scanned to locate free memory of a specified size that is a multiple of the block size. The implementation relies on specific processor instructions related to bit manipulation, and the run-time cost is linear in the worst case. There have been more efficient variants proposed\cite{masmano07}, but these remain fairly complex in terms of implementation complexity.

\end{enumerate}

\subsection{Single Object Memory Allocators}

The most commonly used single object allocation strategy employs the use of buckets. These are lists of memory chunks of a fixed size, bucketed by size. Typically, the buckets are either multiples of a fixed page size or memory blocks that are powers of 2. This strategy results in efficient allocation/deallocation cost, but suffers from the problem of randomizing memory blocks over the lifetime of a long running application, since these free lists are maintained in the order in which memory was returned back to the memory allocator.

This paper builds on prior work which describes a bitmap allocator\cite{dhruv06} that has the desirable memory-locality/cache-friendliness properties, but not the desirable efficient running time in the worst-case.

\section{The Fast Bitmap Fit Data Structure}

\subsection{Advantages relative to linked-free-lists}

\begin{enumerate}

\item \textbf{No need to access user-memory}: The Fast Bitmap Fit allocator maintains metadata separate from the data blocks. A an allocator that uses arenas and free-lists uses an intrusive linked list to chain together free blocks of the same size. This means that a traversal of the linked list necessarily needs to bring in the memory that will be returned to the user into the CPU's cache.

\item \textbf{No memory alignment requirements}: Since the Fast Bitmap Fit allocator maintains a bitmap in a separate part of the memory, the user-memory doesn't need to align memory to pointer-size width (which is necessary in case the free-list pointers need to be maintained in the same location as the memory block to be returned to the user). Even $1$-byte blocks can truly be allocated and returned to the calling application. This means no forced internal fragmentation because no lower-bound is imposed by the allocator itself on the size of each user-returnable memory block.

\end{enumerate}

\subsection{The structure in detail}

Let's take a concrete example. Assume that the allocator pre-allocates space for 8 objects of a fixed size $S$. Hence, we first allocate a memory block of size $S \times 8$ bytes. Each one of these $8$ memory blocks will have an associated bit indicating whether it is available $(0)$ or used $(1)$. These $8$ bits will make up the leaf levels of a complete binary tree-structure. This structure will have $8-1 = 7$ internal nodes, and hence $7$ additional bits for internal nodes. In all, we need $8+7 = 15$ additional bits of data to store metadata about which memory locations are available/used, and how to locate them quickly.

The bitmap is shown graphically below, with bit positions ranging from $0$ to $14$. This bitmap corresponds logically to the full binary tree also shown in the image below.

\textbf{Significance of a bit's value}: The bit at position $0$ is the root node of the entire structure. If this bit has a value $0$, then it means that there is at least one free memory location available in the structure. If this bit has a value $1$, then it means that all the memory location in this structure are occupied.

\textbf{Left/Right Child Arithmetic}: To find the left child of a node at index $\texttt{idx}$, we go to the bit at position $\texttt{idx} \times 2 + 1$, whereas to find the right child of the same node, we visit the bit at position $\texttt{idx} \times 2 + 2$.

\medskip

\includegraphics[width=130mm]{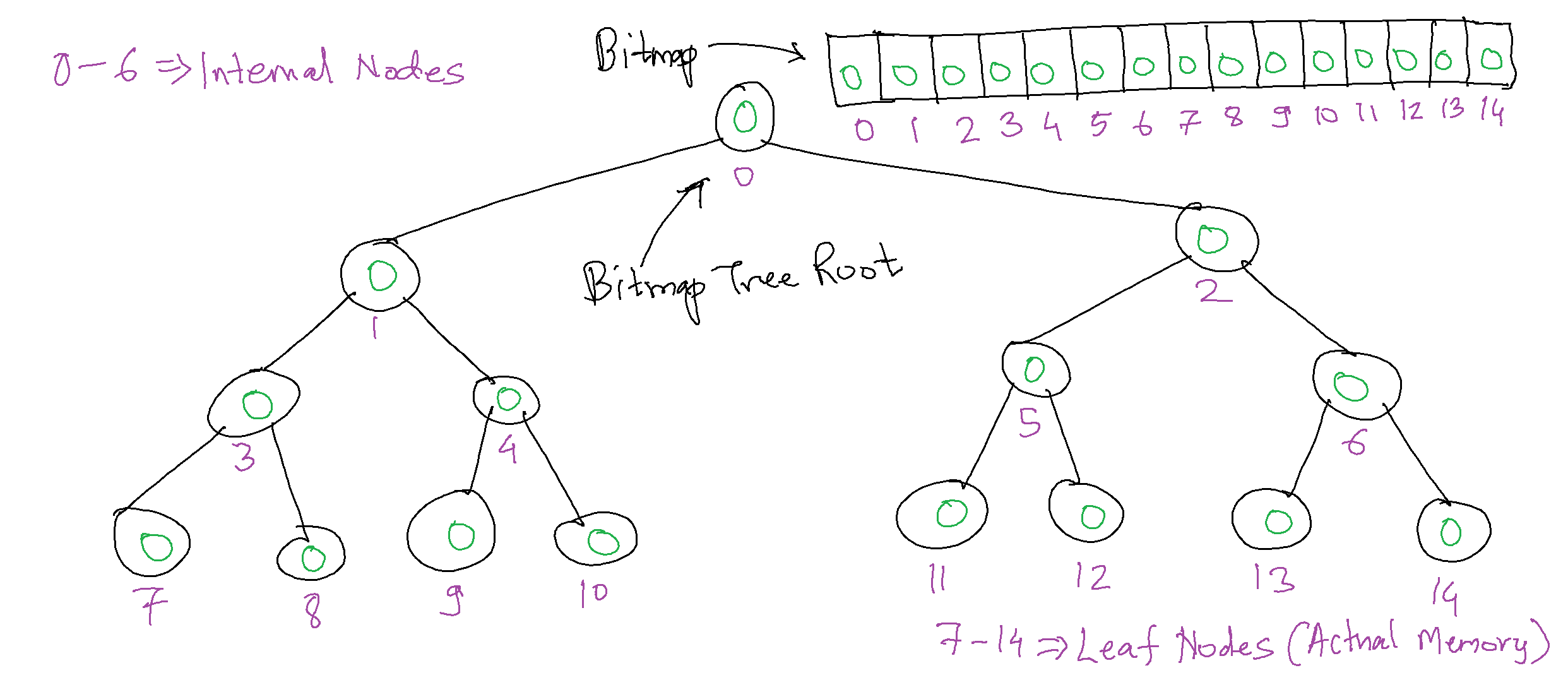}

\subsection{The Allocate Operation}

First, we start at the root (bit-$0$), and walk our way down left or right (preferring the left sub-tree if there's a free object available in the sub-tree) till we reach a leaf node. In the example below, we start at node $0$, and end up at node $7$.

\medskip

\includegraphics[width=130mm]{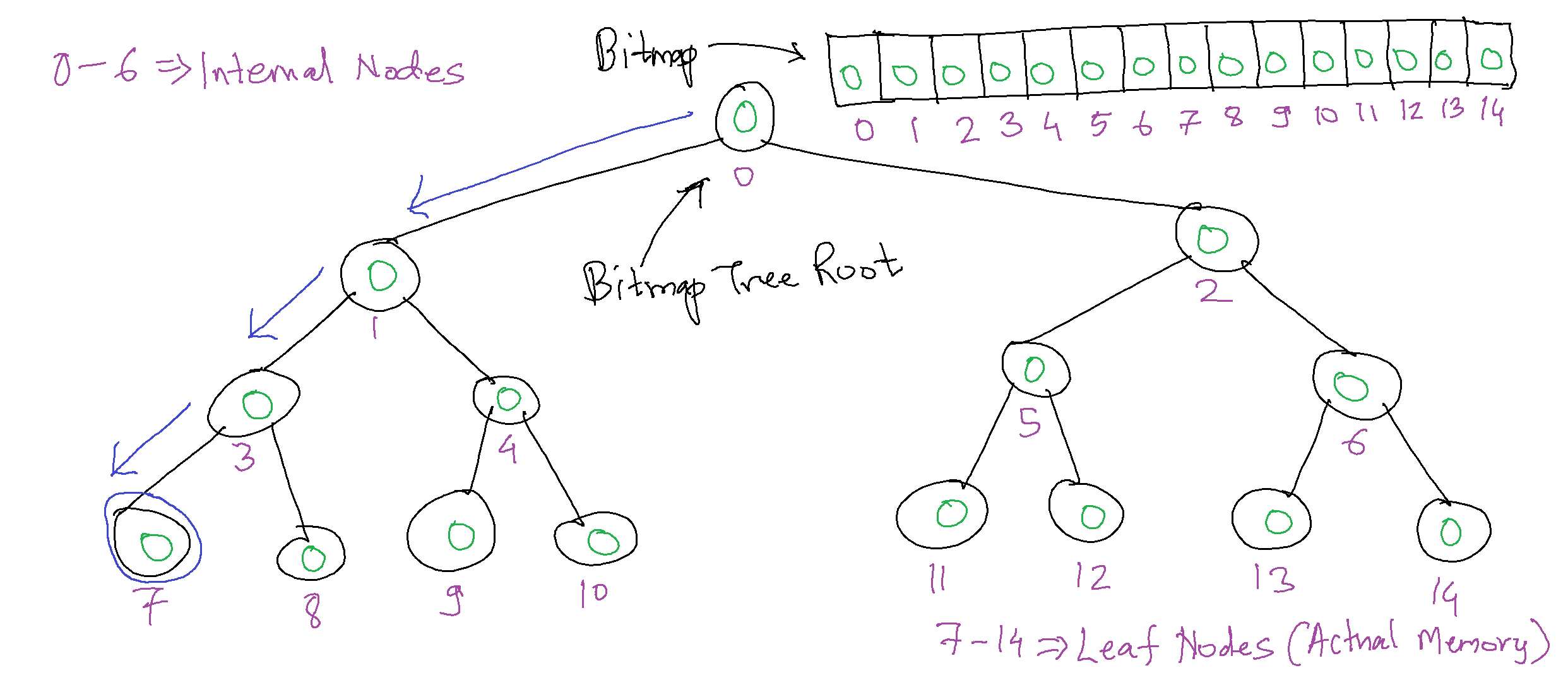}

Now, we mark node $7$ as unavailable by setting the bit at index $7$ to $1$, and walk our way up the tree, always performing the logical $AND$ of sibling nodes to determine the value of the parent node. Once we hit a node where we didn't update the value of the bit in that internal node, we can safely stop propagating up to the root of the tree.

\medskip

\includegraphics[width=130mm]{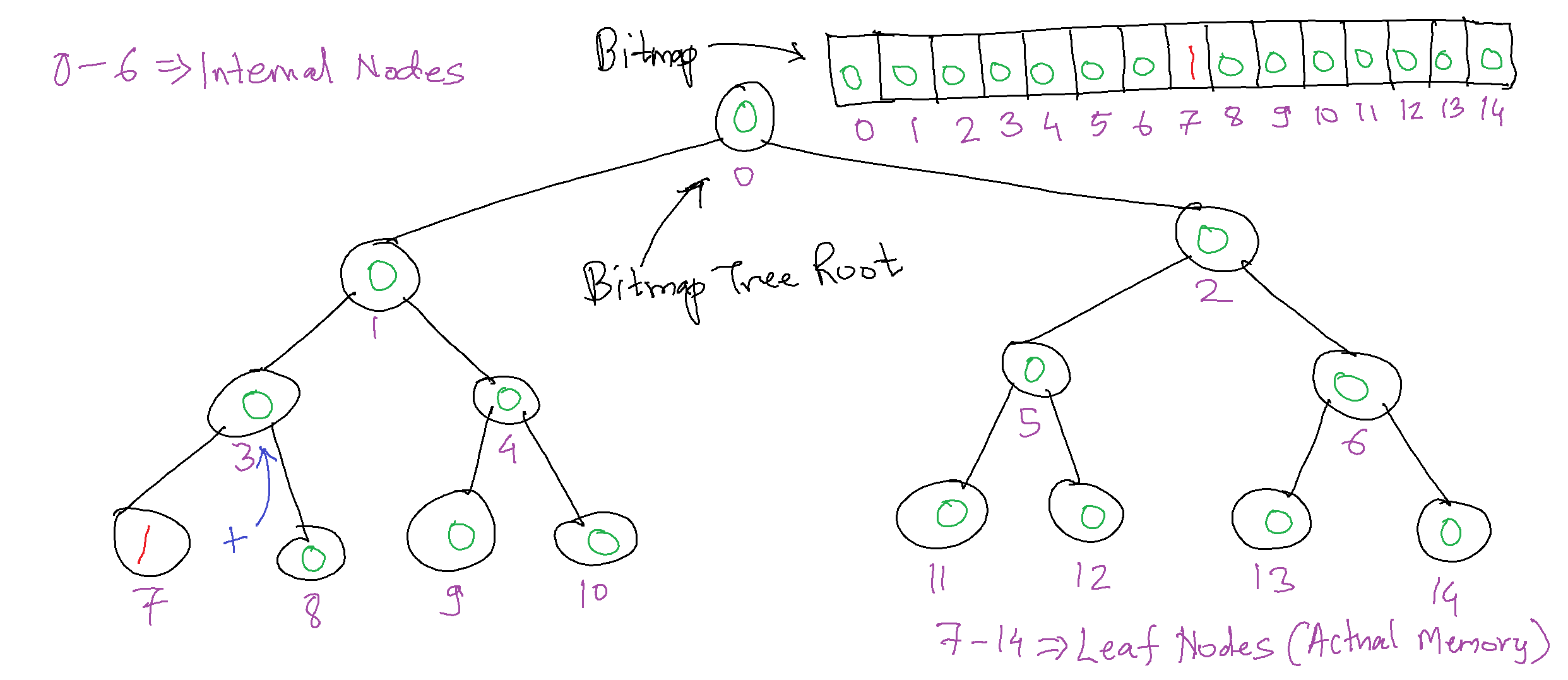}

The graphic below shows one more allocate operation being performed where another memory slot is marked as used.

\medskip

\includegraphics[width=130mm]{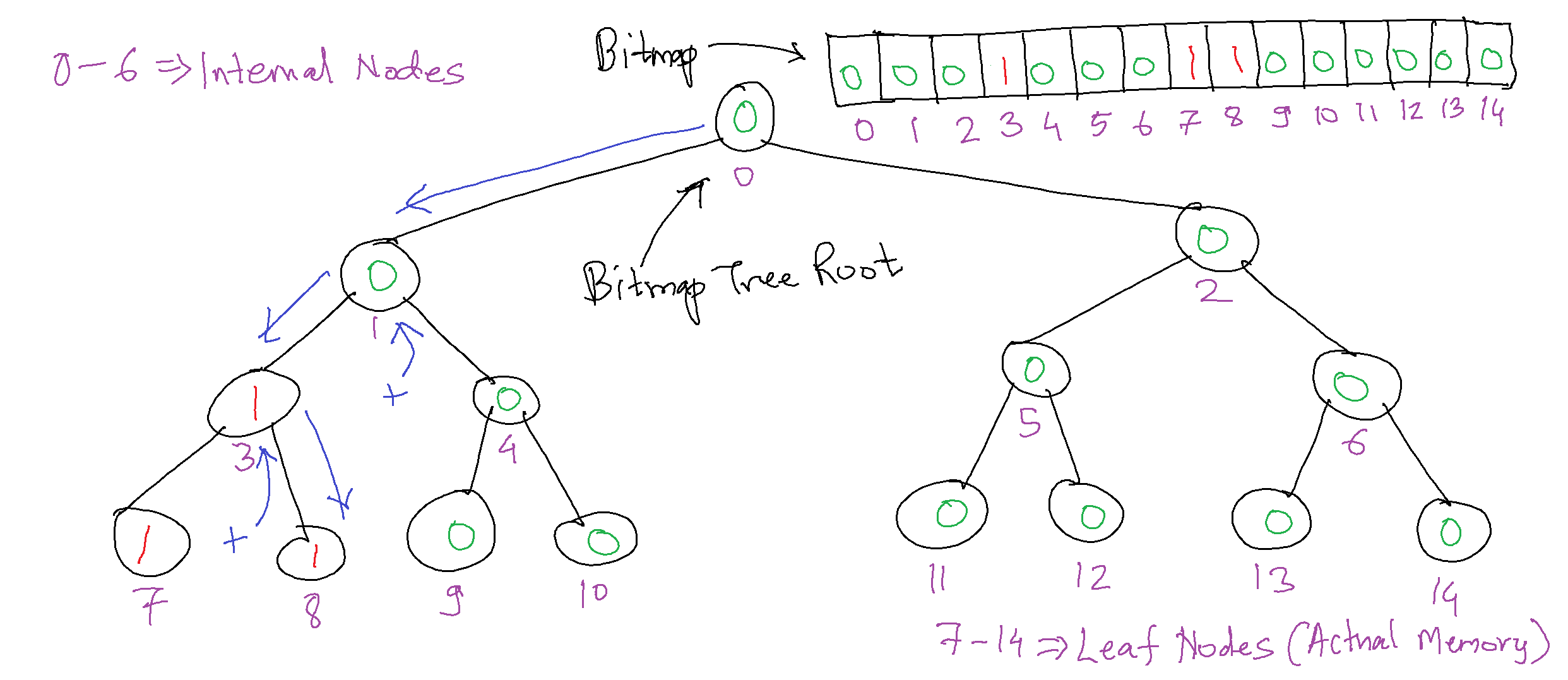}

The cost of each Allocate operation is $\Theta(\log{}n)$.

\subsection{The Free/Deallocate Operation}

Let's assume we're starting with a tree which has $6$ nodes used as shown below.

\medskip

\includegraphics[width=130mm]{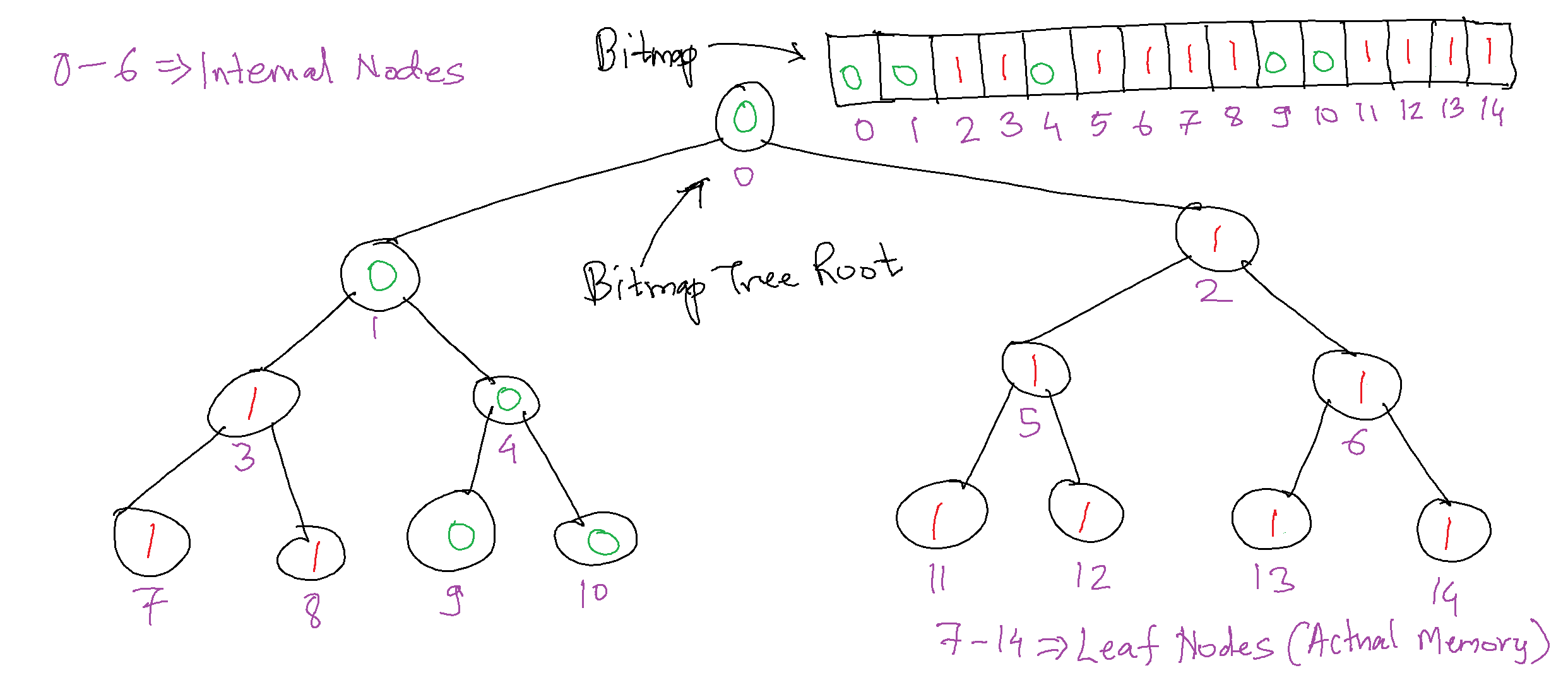}

Suppose the user requested to free the $6^{th}$ memory location (node $12$). We'd start at bit number $12$ in our bitmap, and keep walking up, setting all visited bits to $0$ as we walk up to the root. We can stop as soon as we find a bit that is already $0$ since all bits above it (up to the root) will already be $0$ by definition once we find an internal node with the value $0$.

\medskip

\includegraphics[width=130mm]{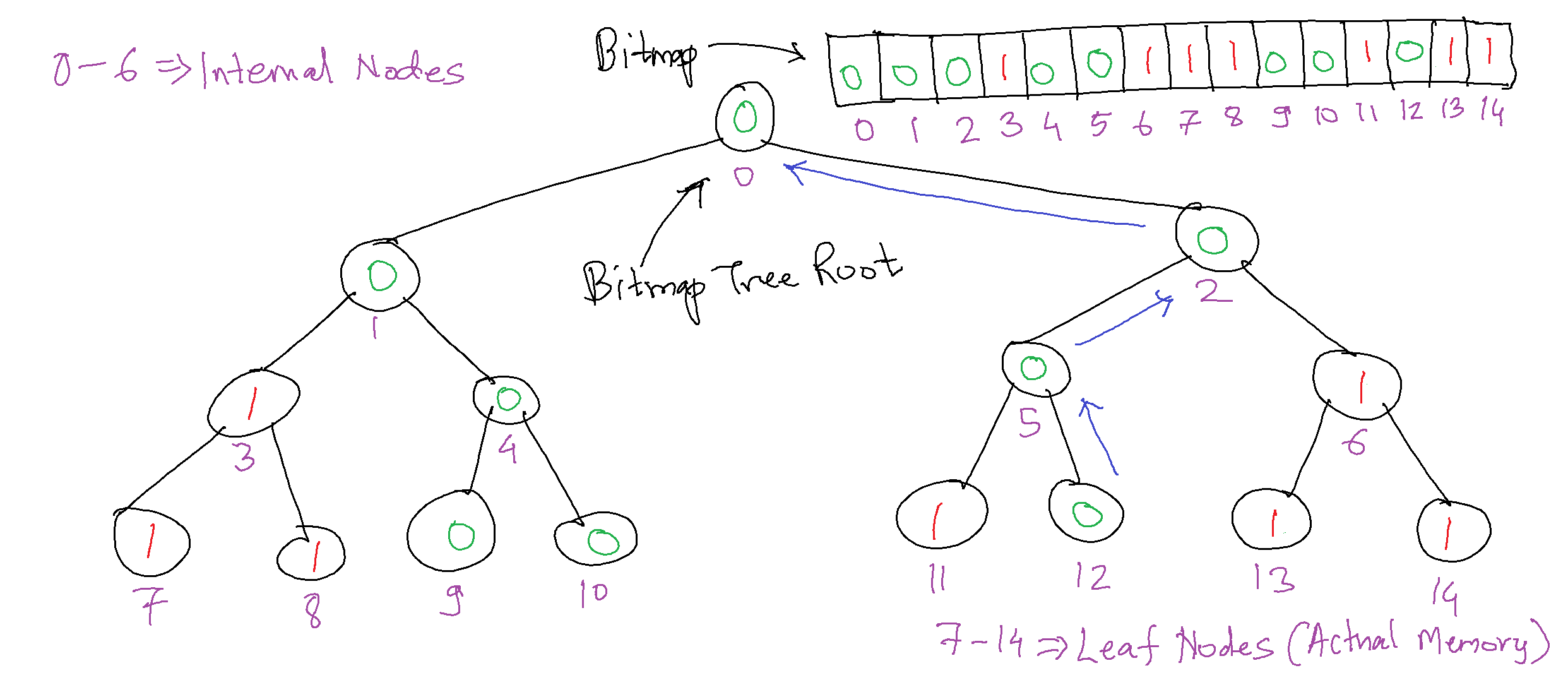}

The cost of each Free/Deallocate operation is $O(\log{}n)$.

\subsection{Allocate with hint}

Some applications may require allocating memory that is close to an already piece of allocated memory because it is expected that both objects will be used simultaneously. In such a scenario, it is beneficial for both the objects to live on the same page that is cached in the CPU's L1/L2 cache. This allocator design supports this operation too to some degree.

The algorithm below isn't guaranteed to find nodes as close to the target node as possible, but assuming a full factor of ~60\%-80\%, there's a good chance that one will find a node next to the provided "hint" node's address.

The general idea is to traverse down from the root of the tree, moving down in the direction of the target (hint) node as long as there is a free node available in that subtree. If there's a choice to be made to go down the left or right subtree, choose the side that has free nodes closer to the "hint" node's location.

\begin{itemize}
\item In the image below, the node $11$ is the "hint" node.
\item We start at the root node $0$ and try to move down the tree toward the node $11$ (right child).
\item When we get to node $2$, we notice that it's occupied, and we back up and instead go to node $1$ (left child of the root), which has a free node under it.
\item Here, we need to choose which side we wish to explore first. We choose the right child of this node (node $4$) since that node has children closer in memory to the hint node (node $11$).
\item Again we move down the right child of node $4$ to node $10$ (which is a leaf node), and stop here.
\end{itemize}

You can see that the node located is close in space (memory) to the hint node.

\medskip

\includegraphics[width=130mm]{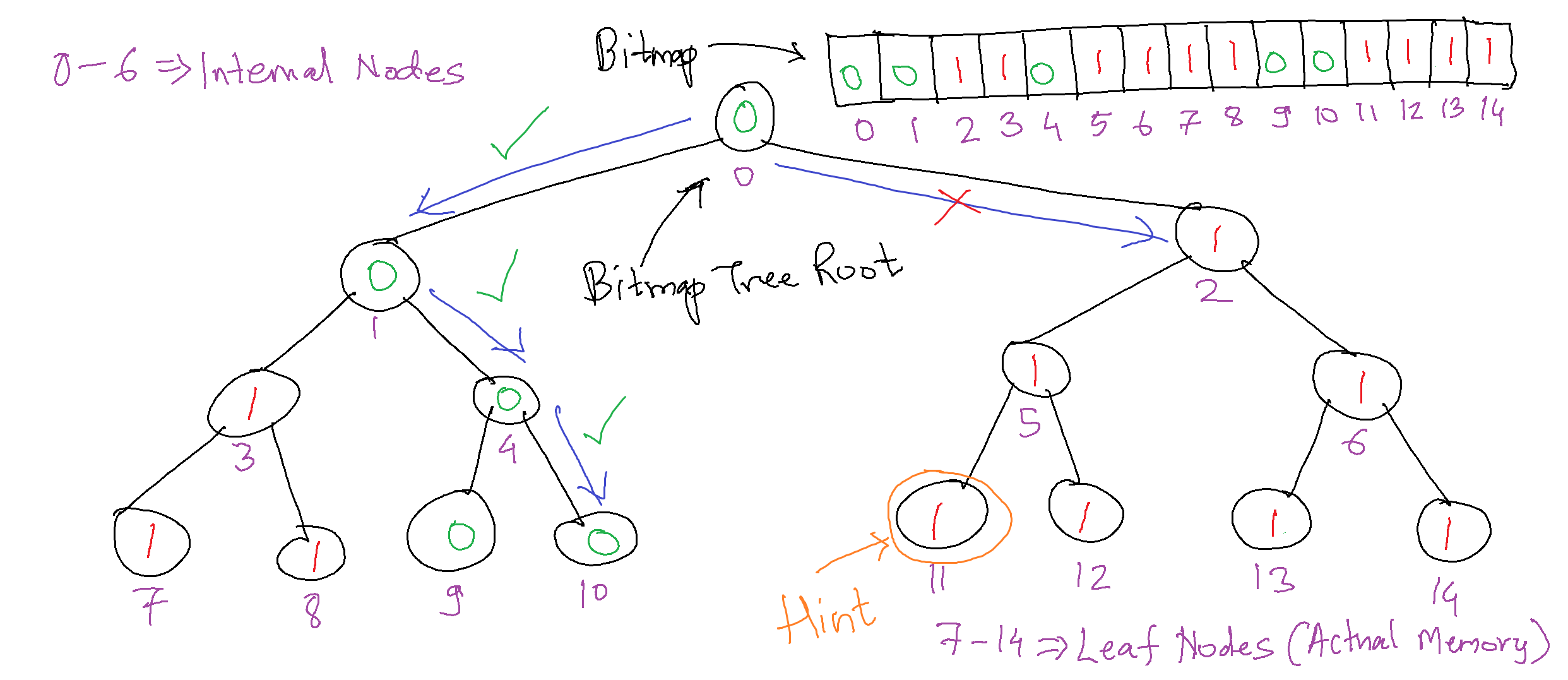}

We now mark nodes as allocated as we did in the allocate method by walking up the parents of the left node, performing a boolean AND of the use/free bits at every level.

\medskip

\includegraphics[width=130mm]{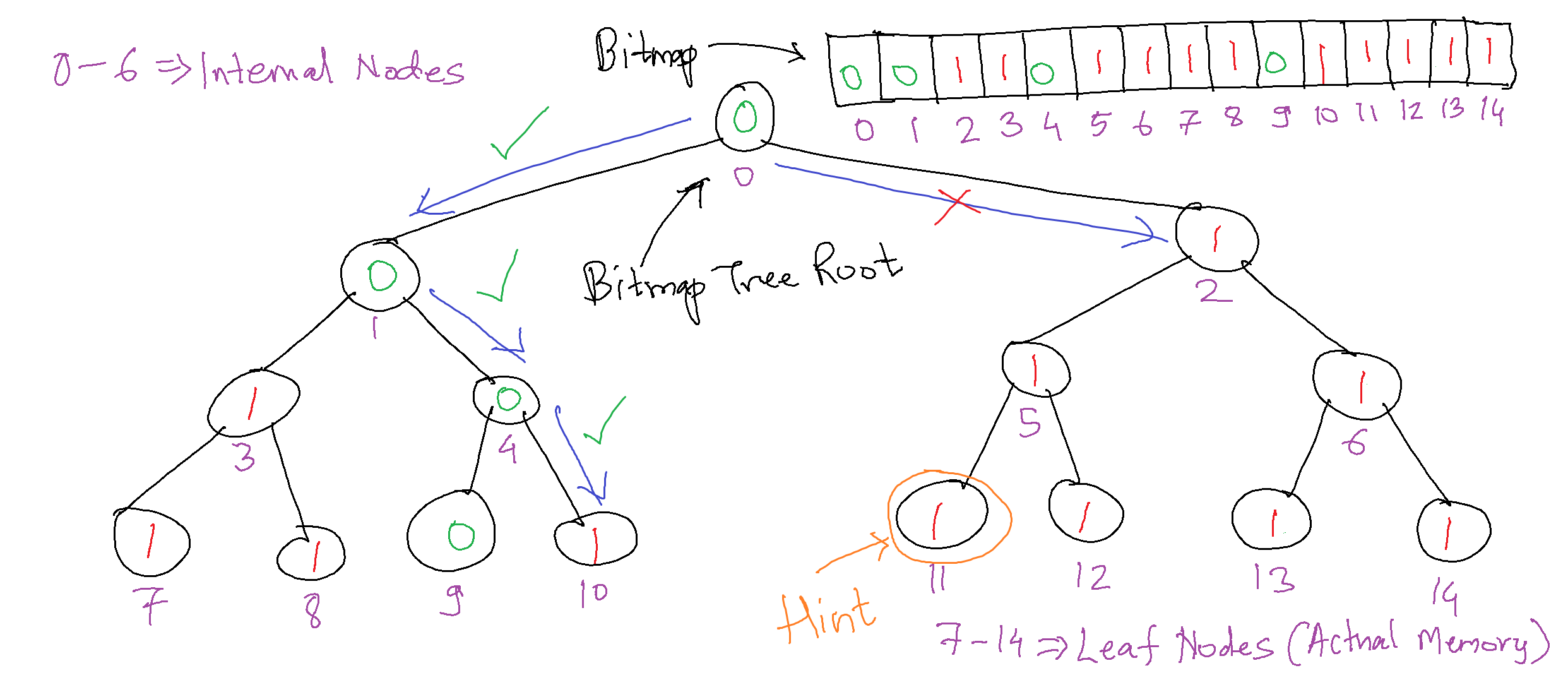}

The cost of each Allocate-with-hint operation is $\Theta(\log{}n)$.

\section{Conclusion}

In this paper, we presented a data structure for a fast bitmap fit memory allocation strategy, which supports allocation (with or without hint) in time $\Theta(\log{}n)$, and free/deallocate in time $O(\log{}n)$. This is a significant improvement over the cost for bitmap fit known so far.

In addition, this approach doesn't mix data with metadata, so sequences of allocate/free operations with no access to the data in between will incur no additional cache activity.

Irrespective of what order the memory is returned back to the allocator, subsequent allocations will always attempt to return objects in memory order to maximize re-use of cached pages in the CPU's L1/L2 caches.

\end{document}